\def\nn#1 #2{#2. #1}				
\def\nnn#1 #2 #3{#2. #3. #1}			
\def\nnnn#1 #2 #3 #4{#2. #3. #4 #1}		
\def\nnnnn#1 #2 #3 #4 #5{#2. #3. #4 #5. #1}	

\def\pp{\noindent\parshape 2 0truecm 13.6truecm 1truecm 12.6truecm}
\def\rl#1;#2;#3;#4;#5 {\addtocounter{enumi}{1}\item[{[\arabic{enumi}]}]\par``#1", #2, {\it #3}, {\bf #4}, #5 \par}
\def\rlnonref#1;#2;#3;#4;#5 {\addtocounter{enumi}{1}\item[{ [\arabic{enumi}]}]\par``#1", #2, {\it #3}, {\bf #4}, #5 \par}
\def\pop#1;#2;#3; {\addtocounter{enumi}{1}\item[{[\roman{enumi}]}]\par``#1", {\it #2}, #3 \par}

\def\rf#1;#2;#3;#4 {\par\pp#1, {\it #2}, {\bf #3}, #4. \par}
\def\rk#1;#2;#3;#4;#5 {\par``#1", #2, {\it #3}, {\bf #4}, #5 \par}

\def\beq#1{\begin{equation}\label{#1}}
\def\eeq{\end{equation}}
\def\beqa#1{\begin{eqnarray}\label{#1}}
\def\eeqa{\end{eqnarray}}

\def\fig#1{Figure~\ref{#1}}
\def\Fig#1{Figure~\ref{#1}}

\def\tabl#1{Table~\ref{#1}}

\def\eg{{\frenchspacing\it e.g.}}
\def\etc{{\frenchspacing\it etc.}}

\def\spose#1{\hbox to 0pt{#1\hss}}
\def\simlt{\mathrel{\spose{\lower 3pt\hbox{$\mathchar"218$}}
    \raise 2.0pt\hbox{$\mathchar"13C$}}}
\def\simgt{\mathrel{\spose{\lower 3pt\hbox{$\mathchar"218$}}
    \raise 2.0pt\hbox{$\mathchar"13E$}}}
\def\simpropto{\mathrel{\spose{\lower 3pt\hbox{$\mathchar"218$}}
    \raise 2.0pt\hbox{$\propto$}}}

\def\d{{$^{\circ}$}}

\documentclass[useAMS,usenatbib,usegraphicx]{mn2e} 
\usepackage{times}

\title[Clustering from 151 MHz to 232 MHz]{Clustering of Extragalactic Sources from~151~MHz to~232~MHz: 
      					   Implications for Cosmological 21-cm Observations}

\author[de Oliveira-Costa \& Lazio]{Ang{\'e}lica de Oliveira-Costa$^1$ \& Joseph Lazio$^{2,3}$ \\
      $^1$Harvard-Smithsonian Center for Astrophysics, Cambridge, MA 02138, USA \\
      $^2$Naval Research Laboratory, Washington, DC  20375, USA \\
      ${}^3$NASA Lunar Science Institute, NASA Ames Research Center,
	Moffett Field, CA  \\
      }

\begin{document}
\input{epsf.sty}

\date{\today. To be submitted to MNRAS.}
\pagerange{\pageref{firstpage}--\pageref{lastpage}}

\pubyear{2010}

\maketitle

\label{firstpage}

\begin{abstract}
In order to construct accurate point sources simulations at the 
frequencies relevant to 21-cm experiments, the angular correlation 
of radio sources must be taken into account. This paper presents 
a measurement of angular two-point correlation function, $w(\theta)$, 
at 232 MHz from the MIYUN survey -- tentative measurements of 
$w(\theta)$ are also performed at 151 MHz. It is found that double 
power law with shape $w(\theta) = A \theta^{-\gamma}$ fits the 
232 MHz data well. For the angular lenght of $0.2^\circ \simlt \theta \simlt
0.6^\circ$, $\gamma \approx -1.12$, and this value of slope is independent 
of the flux-density threshold; while for angular lenghts much greater 
than 0.6\d, $\gamma$ has a shallower value of about $-$0.16. By comparing 
the results of this paper with previous measurements of $w(\theta)$, 
it is discussed how $w(\theta)$ changes with the change of frequency 
and completness limit. 

\end{abstract}



\setcounter{footnote}{0}

\section{Introduction}\label{intro}

There has been considerable recent interest in using the highly
redshifted hyperfine line from H\,\textsc{i} (21-cm line) for
astrophysical and cosmological studies at redshifts $z > 6$
\cite{pl08}. At these redshifts, the 21-cm line is redshifted 
to metre wavelengths, and extracting the cosmological signature 
requires accurate modeling and removal of the foreground Galactic 
and extragalactic emission \cite{santos05,miguel05,wang}.

The diffuse Galactic foreground emission fluctuates mainly 
on large angular scales \cite{angelica}, or on scales that 
are much larger than the expected angular fluctuations of 
the 21-cm signal. Point source contamination from discrete 
extragalactic radio source, on the other hand, affects 
mainly small angular scales and potentially could be more 
problematic.
A number of surveys of radio sources have been performed at 
frequencies relevant to the 21-cm tomography -- see Figure~1 
in \cite{cohen4}; and analysis of these catalogs have helped 
to bring some understanding about their statistical properties: 
for instance, it is known that the distribution of radio 
sources obeys a Poisson statistics with an observed angular 
clustering -- see \tabl{cooresults}\footnote{
	The range of frequencies we consider in this analysis 
	correspond to a redshift range $z \approx 6$--8, 
	which are well within the range being considered for 
	the Epoch of Reionization.}. 

\begin{figure} 
\vskip+0.5cm
\centerline{\epsfxsize=4.2cm\epsffile{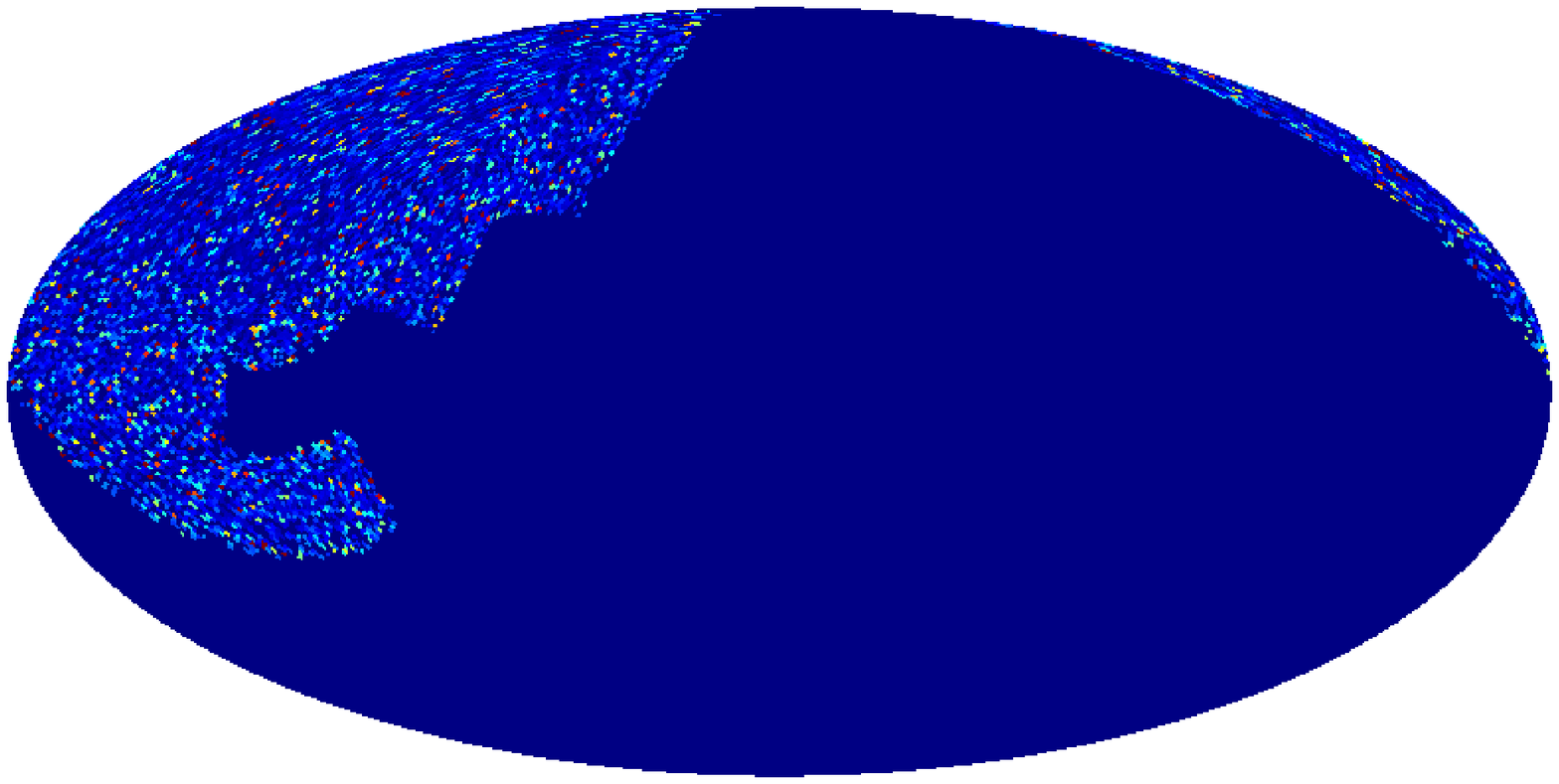}
           \epsfxsize=4.2cm\epsffile{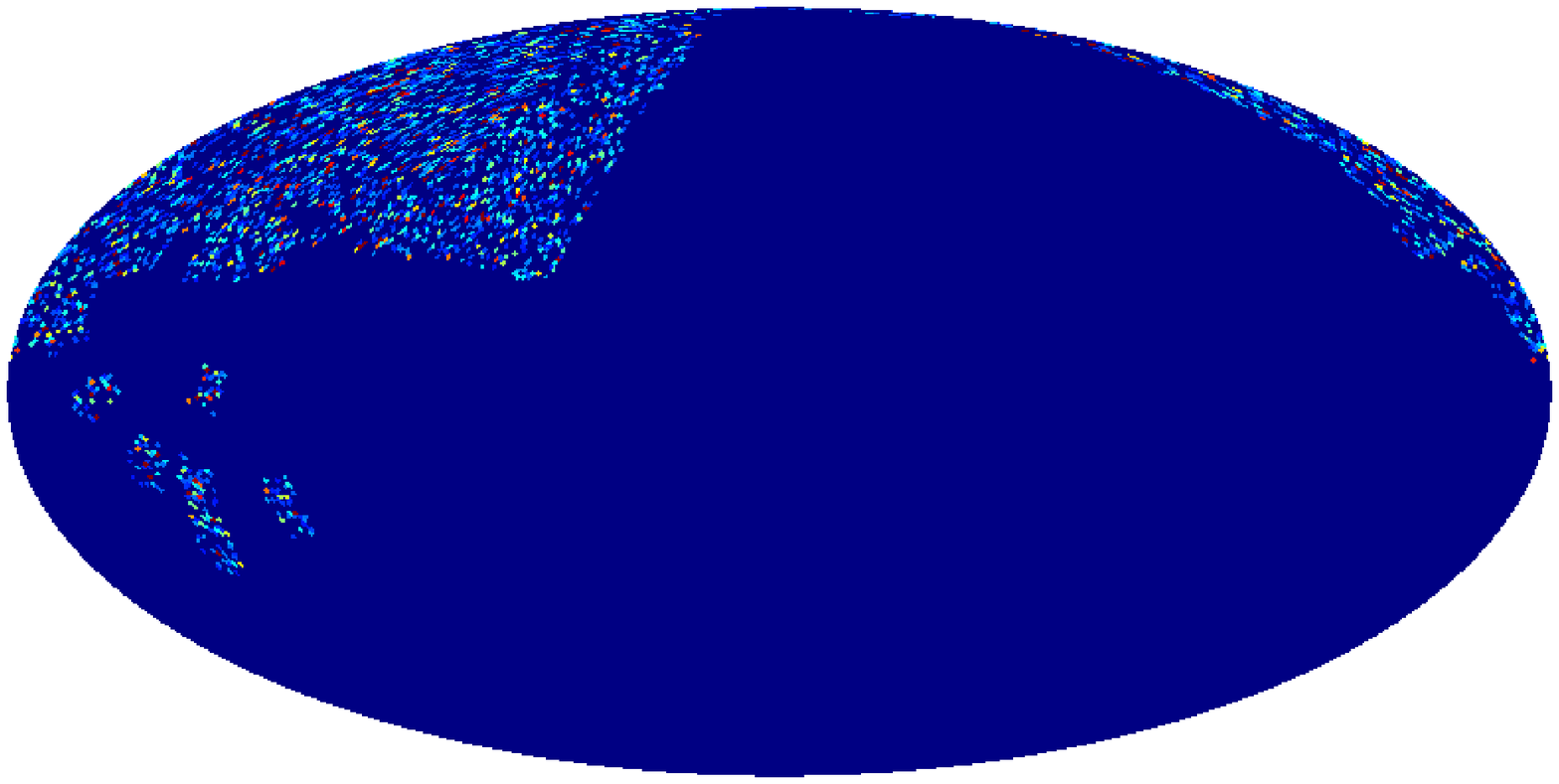}}
\vskip0.2cm
\centerline{\epsfxsize=4.2cm\epsffile{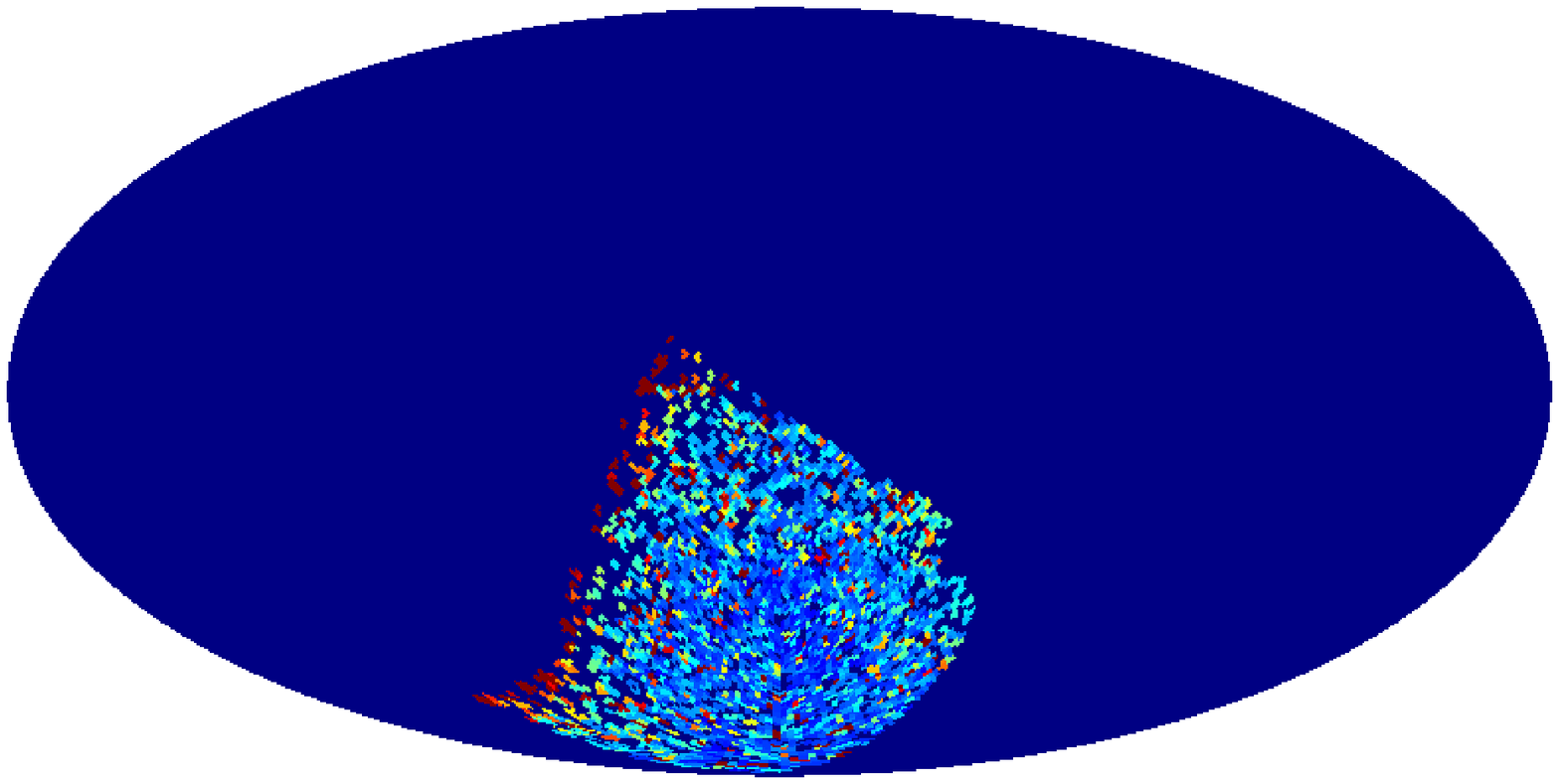}
           \epsfxsize=4.2cm\epsffile{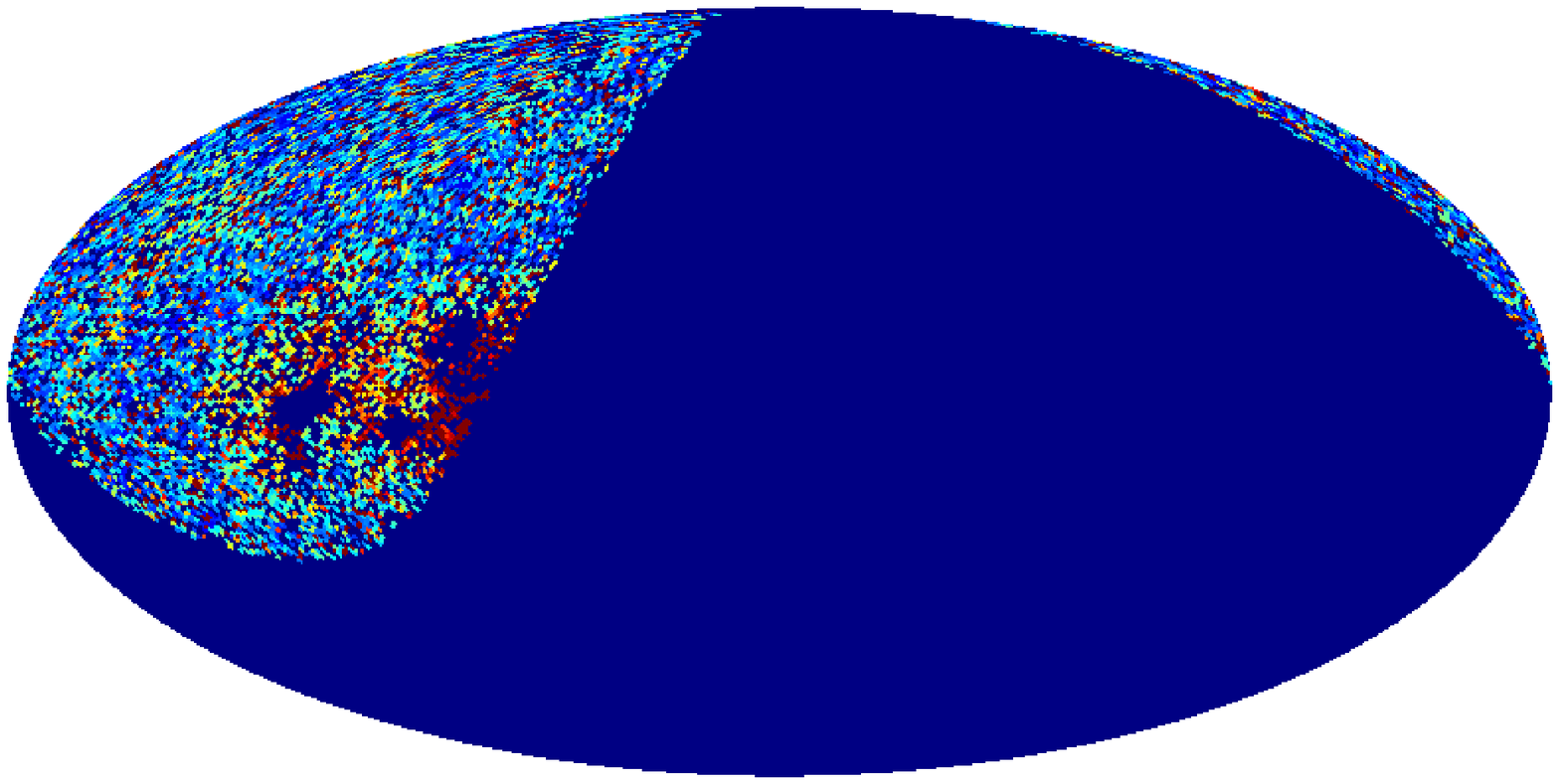}}

\vskip-0.0cm
\caption{
	 Footprints of 
         the 6C        151~MHz catalogue ({\it top,     left}),  
	 the 7C        151~MHz catalogue ({\it top,    right}),  
	 the MRT       151~MHz catalogue ({\it bottom,  left}) and 
	 the MIYUN     232~MHz catalogue ({\it bottom, right}).
	 All footprints were computed using the HEALPix projection 
	 \protect\cite{healpix}, on which locations covered by the 
	 catalogs are in a brighter color (we adopted the convention 
	 in which an increase in flux density~$S$ corresponds to an 
	 increase in color's brightness).
	 All  catalogues are plotted in the interval of 
	 $0 \le S \le 1$ Jy, and in Galactic coordinates 
	 with the Galactic center at the origin and longitude 
	 increasing to the left.  
        } 
\label{4mass} 
\vskip0.2cm
\end{figure}

In this respect, our knowledge of the extragalactic radio source
clustering properties is important because, if there is clustering 
on small angular scales, it could contribute power to a power 
spectral analysis; as a result, the clustering signal could be 
confused with the 21-cm signal that is being sought.
Further, in order to construct accurate simulations at metre 
wavelengths\footnote{
  Some aspects of both experimental design optimization and actual 
  data analysis require full-blown simulations of the sky signal and 
  knowledge about how it propagates through the instrument and the 
  data analysis pipeline. End-to-end simulations are important 
  for 21-cm experiments because of the many complicated issues related 
  to instrumental performance, ionospheric turbulence corrections, 
  {\etc} \protect\cite{miguel03,bowman05,miguel05,judd,judd09}.},
the angular correlation of radio sources must be taken into account 
\cite{gonzalez}. At the faint flux densities relevant for 21-cm
cosmological studies, the relative importance of the clustering
contribution increases and could become an important contribution 
to power spectral analyses.

This paper presents a measurement of angular two-point correlation
function $w(\theta)$ at~232~MHz and a tentative measurement
at~151~MHz. By comparing the results of this paper with previous
measurements of $w(\theta)$, we assess how $w(\theta)$ changes with
the change of frequency and completness limit.
In Section~\ref{tools}, we describe the statistical tools as well as
the surveys used in this analysis. The results and conclusions are
presented in Sections~\ref{results} and~\ref{conclusions},
respectively.


\section{Data Analysis Tools}\label{tools}

\subsection{The Angular 2-point Correlation Function}\label{method}

The clustering of astronomical sources is quantified using 
the angular two-point correlation function~$w(\theta)$. One 
way to estimate this function is to compare the distribution 
of the objects in the real catalogue to the distribution of 
points in a random Poisson distributed catalogue with the 
same boundaries \cite{H93}, or 
\beq{w}
	w(\theta) = \frac{ {\rm DD}(\theta) * {\rm RR}(\theta)}
		         {[{\rm DR}(\theta)]^2} - 1 
\eeq
where DD$(\theta)$, RR$(\theta)$ and DR$(\theta)$ 
are the numbers of data-data, random-random and data-random 
pairs separated by the distance $\theta + \delta\theta$. 

The estimation of RR$(\theta)$ and DR$(\theta)$ requires a catalogue
of objects distributed uniformly over an area with the same angular
boundaries as the data catalogue. In order to produce such catalogues,
we used the ``Sphere Point Picking Algorithm"\footnote{
	http://mathworld.wolfram.com/SpherePointPicking.html.
	}
to generate random cartesian vectors equally distributed on the
surface of a unit sphere (to avoid having vectors ``bunched" around
the poles, as would happen if spherical coordinates were used).
See \cite{gsmpts1} for more details on how these catalogues 
were produced.

Incomplete knowledge of the completeness limit of a survey can affect
our determination of the extent of clustering \cite{magliocchetti2}.
If a completeness limit for a given survey cannot be found in the
literature, we estimate a {\it ballpark} value from the survey's
differential source counts~$dN/dS$.
We derive $dN/dS$ by binning the sources in flux density with bins 
of 0.1 in width, and the bins are not weighted by $S^{-2.5}$.

\begin{figure} 
\vskip-0.3cm
\includegraphics[width=8.7cm]{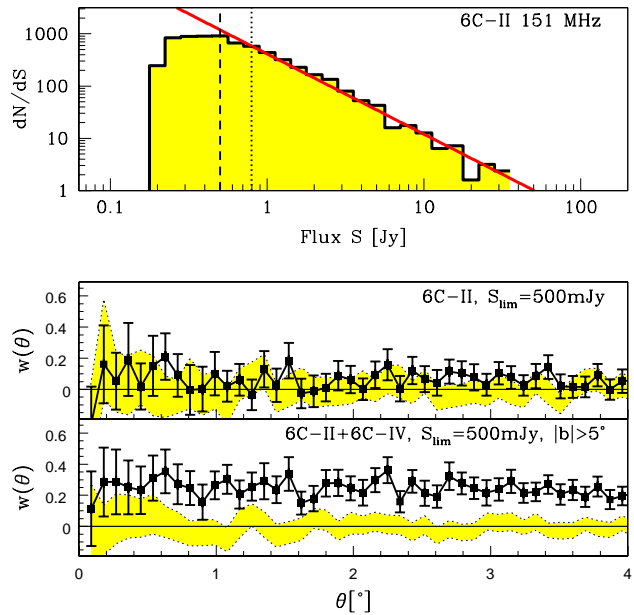}
\caption{The measured source counts~$dN/dS$ ({\it top}) and angular 
        correlation function~$w(\theta)$ ({\it middle}, {\it bottom}) 
	 of the 6C 151~MHz catalogue. 
	 {\it Top:}
	 The source count of the 6C-II region is plotted 
	 using the peak amplitudes. The red line is a
	 single power law fit to the histogram (black line), 
	 for which
	 	$dN/dS = 2.61 - 1.54S$.
	 {\it Middle:}
	 The angular correlation function of the 6C-II region 
	 calculated for a flux density limit of $S = 500$~mJy.
	 {\it Bottom:}
	 The angular correlation function of the (6C-II + 6C-IV) 
	 region calculated for a flux density limit of 
	 $S = 500$~mJy, restricting to Galactic latitudes 
	 $|b| > 5^\circ$. 
	 The yellow shaded regions in the {\it middle} and
	 {\it bottom} panels represent $w(\theta)$ 
	 calculated using solely mock catalogues.
        } 
\label{corr_6c} 
\end{figure}

\subsection{The Catalogues} 

All the catalogues decribed in this subsection are avaiable 
at {\it http://vizier.cfa.harvard.edu/viz-bin/VizieR}, and are
shown in \fig{4mass}. Below we provide a brief description of 
the data sets used in this analysis.


The 6$^{th}$ Cambridge (6C) survey produced a catalogue of 
radio sources at~151~MHz with 34,418 discrete sources at an 
angular resolution of $4.2^\prime \times 4.2^\prime\csc(\delta)$. 
The data product is a set of seven sections (named \hbox{6C-I}, 
\hbox{6C-II}, \hbox{6C-III}, \hbox{6C-IV}, 6C-Va, 6C-Vb, and 
6C-VI) that maps the sky north of $+30^{\circ}$ declination, 
with limiting flux densities between~130~mJy and~200~mJy, 
depending on the section analysed
	\cite{baldwin85,6Cout,hales90,hales91,hales93a,hales93b}.
The uncertainties on the positions and flux densities of many 
sources in the regions 6C-Va, 6C-Vb and 6C-VI are not well 
quantified \cite{hales93a,hales93b}, so we exclude these 
regions from our analysis. 

The 7$^{th}$ Cambridge (7C) survey produced a catalogue of 
radio sources at~151~MHz with 43,683 discrete sources at an 
angular resolution of $1.2^\prime \times 1.2^\prime\csc(\delta)$. 
This survey is composed by combining 96 individual images 
at declinations greater than $+21^{\circ}$. These images have 
completness limits between~120~mJy and~770~mJy, depending on 
the image analysed \cite{7Cout}. 

The Mauritius Radio Telescope (MRT) is a Fourier synthesis 
array that has produced images of the sky covering the region 
	$18^{\mathrm{h}} < \alpha <   24^{\mathrm{h}}$ and 
	$-75^{\circ} < \delta <  -10^{\circ}$ \cite{nayak}. 
The resulting  catalogue contains 2,784 discrete sources at 
an angular  resolution of $4.6^\prime \times 4.6^\prime$ with 
a completness limit\footnote{Shankar 2009, private cominication.} 
of about~1~Jy.


The MIYUN 232~MHz survey mapped the sky north of declination 
$+30^{\circ}$ at an angular resolution of 
$3.8^\prime \times 3.8^\prime\csc(\delta)$, with an average 
noise level of~50~mJy. The principal data product of this survey 
is a catalogue containing 34,426 discrete sources \cite{MIYUN}, 
complete at the 250~mJy level \cite{zhang}.


\begin{figure} 
\vskip-0.5cm
\includegraphics[width=8.7cm]{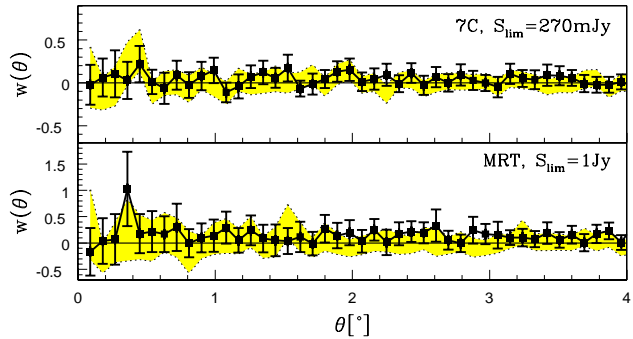}
\vskip-3.8cm
\caption{Measured angular correlation function~$w(\theta)$ of 
	 the 7C ({\it top}) and MRT ({\it bottom}) 151~MHz 
	 catalogues. The angular correlations are calculated 
	 at the flux density limit of $S = 270$~mJy (7C) and 
	 $S = 1$~Jy (MRT). The yellow shaded region is 
	 $w(\theta)$ calculated using solely mocks.
        } 
\label{corr} 
\end{figure}

\section{Results}\label{results} 

\subsection{The catalogues at~151~MHz: \hbox{6C}, \hbox{7C}, \& MRT}\label{151cat}

The completeness limits of the the \hbox{6C-I}, \hbox{6C-II},
\hbox{6C-III}, and 6C-IV regions were not published. Using the 
source counts~$dN/dS$, as described in \S\ref{method}, we estimate
completeness limits of~200~mJy, 500~mJy, 250~mJy, and~500~mJy,
respectively.
\Fig{corr_6c} illustrates our estimation procedure.  The top panel
shows  the differential source count (peak amplitudes) 
of the 8,275 objects in the 6C-II region. The red line corresponds 
to a single power law fit to the $dN/dS$  distribution, for which
	$dN/dS = 2.61 - 1.54 S$.
At flux densities below~800~mJy (vertical black dotted line), the
number counts begin to flatten, and the lack of faint objects becomes
very important below~500~mJy (vertical black dashed line).
We conclude that 
this survey region is incomplete at flux densities below~500~mJy.

\Fig{corr_6c} (middle) shows the measured $w(\theta)$ of the 6C-II 
region for our assumed completeness limit of~500~mJy. The measured 
correlations are represented by the black squares\footnote{
      In order to avoid having Galactic sources in our analysis, we
      adopted Galatic latitude limits and  discarded sources at lower
      Galactic latitudes before measuring $w(\theta)$. 
      100 mock catalogues are constructed using the procedure 
      described in \S\ref{method}, with flux densities above the 
      sensitivity limit of the data catalogue and a chosen 
      Galactic latitude cut (if applicable). 
      By cross-correlating the data with the 100 mocks, a set of
      normally distributed estimates of the correlation function is
      produced. The mean and the standard deviation of this
      distribution are used as a value for the estimate  
      and its uncertainty in the measurement of $w(\theta)$ 
      at each $\theta$. The estimate (mean) and its uncertainty 
      (the standard deviation) are shown as the black squares 
      and their uncertainties (\eg, \fig{corr_6c}, middle). 
      Similarly, the 100 mocks are correlated with themselves.  
      This result corresponds to the yellow shaded region shown 
      in the correlation figures, and, as expected in a Poissonian 
      distribution, $w(\theta)$ is consistent with zero.
      }, 
and distances between data and/or random sources are measured 
in bins of $0.09^\circ$. At this flux density limit, the 6C-II 
catalogue contains 4,111 objects. The results are consistent 
with zero. 
Similar results were obtained for the three remaining regions 
\hbox{6C-I}, \hbox{6C-III}, and 6C-IV.
We also investigated if $w(\theta)$ changes with a change in bin 
size, in Galactic latitude cut, or flux density limit. There is no
indication that modest changes in any of these quantities affect 
the results.

As an extra test, $w(\theta)$ was measured over grouped survey 
regions with similar completness limits. \Fig{corr_6c} (bottom) 
shows $w(\theta)$ calculated for the combined  (6C-II + 6C-IV) 
region, with a completeness limit of~500~mJy and a Galactic 
latitude cut of~$5^\circ$. At this flux density limit, the 
(6C-II + 6C-IV) catalogue contains 6,806 objects, extending 
from~$+30^\circ$ to~$+82^\circ$ in declination. The data show 
the presence of a large scale correlation (which extends 
beyond~$10^\circ$), but no signs of small scale ($\theta < 1^\circ$) 
clustering.
As discussed by \cite{blake02a}, a possible explanation for this
large-scale correlation is a varying areal density of sources on 
the sky, such as might occur from combining two 6C survey regions. 
In the case of 6C-II and~\hbox{6C-IV}, the former covers a 
declination range of~$+50^\circ$ to~$+51^\circ$ while the latter 
covers $+67^\circ$ to~$+82^\circ$.  The change in {\it projected} 
interferometric baselines over this range of declinations would 
produce a varying surface brightness sensitivity, which in turn 
could affect the density of sources on the sky, and thereby 
spuriously enhance the measured value of $w(\theta)$.

\begin{figure} 
\vskip-0.4cm
\includegraphics[width=8.7cm]{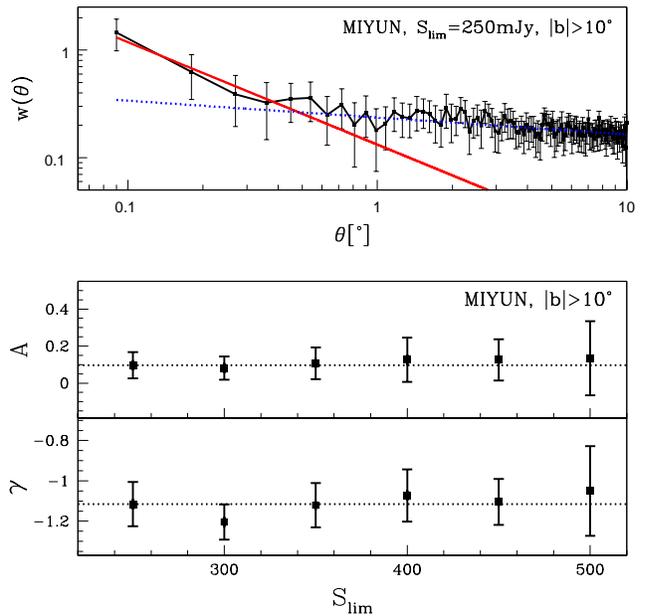}
\caption{The angular correlation function~$w(\theta)$ and 
	 the amplitudes~$A$ and power law indices~$\gamma$ 
	 as a function of flux density determined from the 
	 232~MHz MIYUN catalogue.
	 {\it Top:} The angular correlation function
	 is calculated for a flux density limit of~250~mJy, 
	 for a Galactic latitude cut of~$10^\circ$. 
	 The red solid and the dashed blue lines 
	 are single power laws that fit the data, 
	 with
	 	$w(\theta) = A \theta^{-\gamma}$.
	 {\it Middle} and {\it Bottom:} 	 
	 The amplitudes~$A$ and power law indices~$\gamma$ 
	 are measured for various flux density limits 
	 from~250~mJy to~500~mJy, in steps of~50~mJy. 
	 The amplitude of clustering does not depend 
	 on flux density.
        } 
\label{corr_MIYUN} 
\end{figure}

Like the 6C catalogue, the 7C catalogue was extracted from a survey
composed by multiple individual images that have completeness limits
varying between~120~mJy and~770 mJy \cite{7Cout}. Therefore,$w(\theta)$ 
was measured in each individual region, as well as in aggrouped regions 
having the same completness limit.
\Fig{corr} (top) shows an example of the resulting angular correlation
function for a region with a completeness limit of~270~mJy. At this
flux density limit, the catalogue contains 2,872 objects. Distances
between data and/or random sources are measured in bins
of~$0.09^\circ$. The results are consistent with zero. Similar results
were obtained for the other regions in the 7C catalogue. We also
investigated whether $w(\theta)$ changed with a (modest) change in bin
size or Galactic latitude cut, and there is no indication that any of
these changes affect the results.

\begin{table*}  
\caption{Published w$(\theta)^1$ values.} 
\medskip
\centerline{\label{cooresults}
\begin{tabular}{l|c|c|c|c|r}
\hline
\hline
\multicolumn{1}{ l|}{Ref}	     &
\multicolumn{1}{|c|}{$\nu$}	     &  
\multicolumn{1}{|c|}{$A$} 	     &
\multicolumn{1}{|c|}{$\gamma$}       &
\multicolumn{1}{|c|}{w$(\theta)$}    &
\multicolumn{1}{|r}{S$_{\mathrm{lim}}$} \\
&[GHz] & & &[\d]&[mJy]              \\ 
\hline
\hline
\protect\cite{gsmpts1} 			&0.074	 & 0.103$\pm$0.026  &1.21 $\pm$0.35  &0.2--0.6     & 770     \\
 This Work (6C)		       	 	&0.151   &		    &		     &--	   & 500     \\
 This Work (7C)		        	&0.151   &		    &		     &--	   &120--770 \\
 This Work (MRT)		        &0.151   &		    &		     &--	   &1000     \\
 This Work (MIYUN, $|b|>10^\circ$)	&0.232   & 0.096$\pm$ 0.071 &1.12 $\pm$0.11  &$<$ 0.6	   & 250     \\
\hline
& &${\rm A} \times 10^{-3}$ &&& \\ 
\hline
\protect\cite{seldner}  		&0.178	 & 		    &		     &1.50--3.0    &3000     \\
\protect\cite{Rengelink98}		&0.325	 & 2.0  $\pm$ 0.5   &0.8             &$<$ 1.0	   & 35      \\
\protect\cite{blake}          		&0.325	 & 1.01 $\pm$ 0.35  &1.22 $\pm$0.33  &$>$ 0.2	   & 35      \\
\protect\cite{webster77}                &0.408	 & 		    &		     &--	   & 250     \\
\protect\cite{webster77b}               &0.408	 & 		    &		     &--	   &  10     \\
\protect\cite{blake}          		&0.843	 & 2.04 $\pm$ 0.38  &1.24 $\pm$0.16  &$>$ 0.2	   & 10      \\
\protect\cite{cress}       	        &1.400	 & 3.7  $\pm$ 0.3   &1.06 $\pm$0.03  &0.02--2.0    &   3     \\
\protect\cite{magliocchetti}       	&1.400	 & 2.68 $\pm$ 0.07  &1.52 $\pm$0.06  &0.30--3.0    &   3     \\
\protect\cite{overzier}       		&1.400	 & 1.2  $\pm$ 0.1   &1.8	     &$>$ 0.3	   &   3     \\
\protect\cite{blake}          		&1.400   & 1.49 $\pm$ 0.15  &1.05 $\pm$0.10  &$>$ 0.3	   & 10      \\
\protect\cite{overzier}       		&1.400	 & 1.0  $\pm$ 0.2   &1.8	     &$>$ 0.3	   &  10     \\
\protect\cite{webster77}                &2.700	 & 		    &		     &-- 	   & 350     \\
\protect\cite{sicotte}                  &4.850   & 		    &		     &0.70--1.7    &  45     \\
\protect\cite{kooiman}                  &4.850   & 4.01 	    &0.8 	     &0.30--1.9    &  35     \\
\protect\cite{Rengelink98}		&4.850	 & 6.5  $\pm$ 2.0   &0.8             &$<$ 2.5	   & 35      \\
\protect\cite{loan}                     &4.850   &10.0  $\pm$ 5.0   &0.8 	     &$<$ 2.0	   &  50     \\
\hline
\hline
\end{tabular} 
}
\smallskip
\noindent{\small \\
		 $^1$w$(\theta)$ is fitted by a power-law of the form $A\theta^{-\gamma}$}\\
		 $S_{\mathrm{lim}}$ is the limiting flux density.
\end{table*}

\Fig{corr} (bottom) shows the measurement of $w(\theta)$ 
for the MRT catalogue, with a flux density limit of~1~Jy 
(black squares). At this flux limit, the catalogue 
contains 2,294 objects. Distances between data and/or 
random sources are measured in bins of $0.09^\circ$. 
The results are consistent with zero. We also investigated 
whether $w(\theta)$ changed with a (modest) change in bin 
size, flux density limit, or Galactic latitude cut, and 
there is no indication that any of these changes affect 
the results.

\vskip-1cm
\subsection{The MIYUN catalogue} 

\fig{corr_MIYUN} (top) shows the determination of 
$w(\theta)$ for the flux density limit of~250~mJy
(black squares). Distances between data and/or 
random sources are measured in bins of~$0.09^\circ$. 
It  is apparent that the angular correlation function cannot be fit
with a single  power law, but requires two power laws, each of the form
	$w(\theta) = A \theta^{-\gamma}$ 
\cite{Peebles80}, where $A$ is a measure of the amplitude 
of the average enhancement of the number of radio sources 
at a particular point in the sky.
Fitting the data with a double power law model yields 
a weak correlation with  
$A = 0.096 \pm 0.071$ and $\gamma = -1.12 \pm 0.11$ 	 
(red solid line, with $\chi^2$=0.01), and
$A = 0.236 \pm 0.092$ and $\gamma = -0.16 \pm 0.05$ 
(blue dotted line, with $\chi^2$=0.24).
The  break in the angular correlation function occurs
between $0.4^\circ \simlt \theta \simlt 0.6^\circ$. 
This break could be the indication that there is a 
small angular scale signal (maybe caused by clustering) 
added to a large angular scale signal in the 
survey\footnote{ 
	As pointed out by \protect\cite{blake02a}, 
	in a wide-area survey undertaken with an 
	interferometer (such as the MIYUN), large 
	scale gradients in completeness limit 
	typically appear as a function of declination.
	}.  

We  also investigated whether $w(\theta)$ changed with 
a (modest) change in bin size or Galactic latitude cut.  
Variations in the bin size did not affect the
results, but  for Galactic latitude cuts greater than 
$+15^\circ$,  $w(\theta)$ is consistent with zero.  

The angular correlation function~$w(\theta)$ was also calculated for
various flux density 
limits from~300 to~800~mJy, in increments of~50~mJy.  Above
the  flux density limit of~500~mJy, the correlation for small angular
scales ($\theta < 0.6^\circ$) approaches zero. As shown in 
\fig{corr_MIYUN} ({\it middle} and {\it bottom}), for a $10^\circ$
Galactic latitude cut, the amplitude of clustering does not depend on
flux density. This same result was observed in previous angular
correlation analysis, \eg, \cite{blake,gsmpts1}.


\section{Conclusions}\label{conclusions}

We have used existing 151~MHz and~232~MHz catalogues to estimate the
angular correlation function~$w(\theta)$ relevant for both simulation
and analysis of 21-cm cosmological observations.
At~151~MHz (corresponding to $z \approx 8.4$), our estimate
of~$w(\theta)$ is determined from the \hbox{6C}, \hbox{7C}, and MRT
surveys.  In all cases, the results are consistent with zero, implying
no observed clustering of radio sources, at flux density limits
ranging from about~0.3~Jy to~1~Jy.
At~232~MHz (corresponding to $z \approx 5$), we found that $w(\theta)$
can be fit by a  two broken power laws, each with shape 
	$w(\theta) = A \theta^{-\gamma}$, 
with a break at an angular scale of $\theta \sim 0.5^\circ$.  At small
angular scales, with  a Galactic latitude cut of $10^\circ$ and a flux
density 
limit of~250~mJy, we find
	$A = 0.096 \pm 0.071$ and $\gamma = -1.12 \pm 0.11$ 
(with $\chi^2$=0.01), while, at large angular scales, we find
	$A = 0.236 \pm 0.092$ and $\gamma = -0.16 \pm 0.05$ 
(with $\chi^2$=0.24). The value of~$\gamma$ at small angular 
scales is consistent with that measured from radio catalogues 
at higher frequencies (\tabl{cooresults}).

From \tabl{cooresults}, it appears that the power law index~$\gamma$
of the angular correlation function is essentially constant over the
frequency range from~74~MHz to~232~MHz ($20 \simlt z \simlt 5$), and 
it is also consistent with that determined from higher frequencies, 
at least as high as 1400~MHz.  
Strikingly, however, it is not clear that the angular scales for which 
this power law index applies are consistent across the range of frequencies:  
at the lower frequencies, clustering seems to be important on angular 
scales of $\theta \simlt 0.6^\circ$, while at higher frequencies, the 
clustering is important on angular scales $\theta \simgt 0.3^\circ$. 
These are overlapping ranges, but not entirely consistent.

It is importatnt to point out that the classes of sources being probed 
at the different frequencies are also not entirely the same. Source 
counts at~1400~MHz indicate that strong radio sources (\eg., FR~II 
radio galaxies) dominate at flux densities above about~10~mJy, 
while star forming galaxies become important at lower frequencies.
The current flux density limits for surveys at frequencies
around~200~MHz (and below) are about~500~mJy. Assuming a nominal 
spectral index of 
	$\alpha = -0.7$ ($S_\nu \propto \nu^\alpha$), 
these flux density limits imply flux densities of about~125~mJy 
at~1400~MHz, which is well above the flux density limits probed
by the surveys around~1000~MHz, and consistent with the notion 
that the current generation of low-frequency surveys are dominated 
by powerful radio sources. 
Consequently, one potential explanation is that the possible 
inconsistencies are due, in part, to the different source populations 
being probed.  However, it is also notable that the clustering 
amplitude~$A$ is fairly marginal (always less than 4$\sigma$, 
and in some cases consistent with zero) at frequencies lower 
than 1400~MHz; while at the higher frequencies, by contrast, the
clustering amplitudes can exceed a significance of~10$\sigma$.

For the purposes of 21-cm cosmological observations, either 
simulating sky models or analyzing low radio frequency observations, 
we conclude that it is acceptable to scale the angular clustering 
results from higher to lower frequencies. One caveat to this conclusion 
is if a strongly clustered, steep-spectrum population of objects exists.
Using a fiducial 1400~MHz flux density of~10~mJy allows us to estimate 
how deep future low-frequency surveys might need to be in order to 
assess this possible inconsistency (or the existence of a possible 
steep-spectrum population) in the scale of angular clustering. With 
the nominal spectral index of~$-0.7$, we estimate that a survey 
around~150~MHz\footnote{
	Such as the proposed Million Source Shallow Survey (MSSS) with
	the Low Frequency Array (LOFAR) -- more information at
	http://www.astron.nl/general/lofar/lofar.}
would need to reach a limiting flux density of about~50~mJy (implying
a thermal noise limit of about~7~mJy).

\bigskip
\noindent
{\bf ACKNOWLEDGMENTS:}~~ 
We thank N.~Kassim for illuminating discussions. 
Support for this work was provided by the NSF 
through grants AST-0607597 and AST-0908950. 
The LUNAR consortium is funded by the NASA 
Lunar Science Institute (via Cooperative Agreement 
NNA09DB30A) to investigate concepts for astrophysical 
observatories on the Moon. Basic research in radio 
astronomy at NRL is supported by 6.1 Base funding.




\begin{thebibliography}{} 

\bibitem[Baldwin et al. 1985]{baldwin85}  
        \hspace{-.1in} Baldwin J.E., Boysen R.C., Hales S.E.G., Jennings J.E., Waggett P.C., Warner P.J., Wilson D.M.A., 1985, MNRAS, 217, 717 

\bibitem[Blake et al. 2004]{blake} 
	\hspace{-.1in} Blake C., Mauch T., Sadler E.M., 2004, MNRAS, 347, 787 

\bibitem[Blake \& Wall 2002]{blake02a} 
	\hspace{-.1in} Blake C., Wall J., 2002a, MNRAS, 337, 993 

\bibitem[Bowman 2007]{judd} 
        \hspace{-.1in} Bowman J.D., 2007, ``Probing the Epoch of Reionization with Redshifted 21-cm HI Emission", Ph.D. Thesis, Massachusetts Institute of Technology    

\bibitem[Bowman et al. 2006]{bowman05}
        \hspace{-.1in} Bowman J.D., Morales M.F., Hewitt J.H., 2006, ApJ, 638, 20 

\bibitem[Bowman et al. 2009]{judd09}
        \hspace{-.1in} Bowman J.D., Morales M.F., Hewitt J.H., 2009, ApJ, 695, 18 

\bibitem[Cohen et al. 2003]{cohen4}
        \hspace{-.1in} Cohen A.S., Rottgering H.J.A., Kassim N.E., Cotton W.D., Perley R.A., Wilman R., Best P., Pierre M., Birkinshaw M., Bremer M., Zanichelli A., 2003, ApJ, 591, 640 

\bibitem[Cress et al. 1996]{cress} 
        \hspace{-.1in} Cress C.M., Helfand D.J., Becker R.H., Gregg M.D., White R.L., 1996, ApJ, 473, 7

\bibitem[de Oliveira-Costa \& Capodilupo 2009]{gsmpts1}
        \hspace{-.1in} de Oliveira-Costa A., Capodilupo J., 2009,  arXiv:0908.4248 

\bibitem[de Oliveira-Costa et al. 2008]{angelica}
        \hspace{-.1in} de Oliveira-Costa A., Tegmark M., Gaensler B.M., Jonas J., Landecker T.L., Reich P., 2008, MNRAS, 388, 247 

\bibitem[Gonzalez-Nuevo et al. 2005]{gonzalez}
        \hspace{-.1in} Gonzalez-Nuevo J., Toffolatti L., Argueso F., 2005, ApJ, 621, 1

\bibitem[Gorski et al. 2005]{healpix}
        \hspace{-.1in} Gorski, K.M., Hivon, E., Banday, A.J., Wandelt, B.D., Hansen, F.K., Reinecke, M., Bartelmann, M., 2005, ApJ, 622, 759

\bibitem[Hales et al. 1988]{6Cout}  
        \hspace{-.1in} Hales S.E.G , Baldwin J.E., Warner P.J., 1988, MNRAS, 234, 919 		    

\bibitem[Hales et al. 1993b]{hales93b} 
        \hspace{-.1in} Hales S.E.G., Baldwin J.E., Warner P.J., 1993b, MNRAS, 263, 25

\bibitem[Hales et al. 1990]{hales90}  
        \hspace{-.1in} Hales S.E.G., Masson C.R., Warner P.J., Baldwin J.E., 1990, MNRAS, 246, 256 

\bibitem[Hales et al. 1993a]{hales93a}   
        \hspace{-.1in} Hales S.E.G., Masson C.R., Warner P.J., Baldwin J.E., Green D.A., 1993a, MNRAS, 262, 1057

\bibitem[Hales et al. 1991]{hales91}  
        \hspace{-.1in} Hales S.E.G., Mayer C.J., Warner P.J., Baldwin J.E., 1991, MNRAS, 251, 46 

\bibitem[Hales et al. 2007]{7Cout} 
        \hspace{-.1in} Hales S.E.G , Riley J.M., Waldram E.M., Warner P.J., Baldwin J.E., 2007, MNRAS, 382, 1639

\bibitem[Hamilton 1993]{H93}
        \hspace{-.1in} Hamilton A.J.S., 1993, ApJ, 417, 19 

\bibitem[Kooiman et al. 1995]{kooiman}  
        \hspace{-.1in} Kooiman B.L., Burns J.O., Klypin A.A., 1995, ApJ, 448, 500 

\bibitem[Loan et al. 1997]{loan}  
        \hspace{-.1in} Loan A.J., Wall J., Lahav O., 1997, MNRAS, 286, 994 

\bibitem[Magliocchetti et al. 1998]{magliocchetti2}  
        \hspace{-.1in} Magliocchetti M., Maddox S.J., Lahav O., Wall J., 1998, MNRAS, 300, 257

\bibitem[Magliocchetti et al. 1999]{magliocchetti}  
        \hbox{Magliocchetti M., Maddox S.J., Lahav O., Wall J., 1999, MNRAS,} 306, 943

\bibitem[Morales et al. 2005]{miguel05}
        \hspace{-.1in} Morales M.F., Bowman J.D., Hewitt J.H., 2005, A\&AS, 207, 3304 

\bibitem[Morales \& Hewitt 2004]{miguel03}
        \hspace{-.1in} Morales M.F., Hewitt J.H., 2004, ApJ, 615, 7 

\bibitem[Nayak et al. 2009]{nayak}  
        \hspace{-.1in} Nayak A., Daiboo S., Shankar N.U., 2009, ASPC, 407, 426

\bibitem[Overzier et al. 2003]{overzier} 
        \hspace{-.1in} Overzier R.A., Rottgering H.J.A., Rengelink R.B., Wilman R.J., 2003, A\&A, 405, 53

\bibitem[Peebles 1980]{Peebles80}
        \hspace{-.1in} Peebles P.J., 1980, ``The Large Scale Structure of the Universe", Princeton University Press

\bibitem[Pritchard \& Loeb 2008]{pl08}  
	\hspace{-.1in} Pritchard, J.R., \& Loeb, A.  2008, PhRvD, 78, 103511
	
\bibitem[Rengelink \& Rottgering 1999]{Rengelink98}  
        \hspace{-.1in} Rengelink R., Rottgering H., 1999,  Proc. of the ``The Most Distant Radio Galaxies", Ed. Rottgering H.J.A., Best P.N., Lehnert M.D., Amsterdam, 15-17 October 1997, p. 399

\bibitem[Santos et al. 2005]{santos05}
        \hspace{-.1in} Santos M.G., Cooray A., Knox L., 2005, ApJ, 625, 575 

\bibitem[Seldner \& Peebles 1981]{seldner}  
        \hspace{-.1in} Seldner M., Peebles P.J.E., 1981, MNRAS, 194, 251 

\bibitem[Sicotte \& Peebles 1995]{sicotte}   
        \hspace{-.1in} Sicotte H., Peebles P.J.E., 1995, AIPC, 336, 390

\bibitem[Wang et al. 2006]{wang}
        \hspace{-.1in} Wang X., Tegmark M., Santos M.G., Knox L., 2006, ApJ, 650, 529 

\bibitem[Webster 1977a]{webster77} 
        \hspace{-.1in} Webster A., 1977a, MNRAS, 179, 511

\bibitem[Webster 1977b]{webster77b} 
        \hspace{-.1in} Webster A., 1977b, MNRAS, 179, 517 

\bibitem[Zhang 1999]{zhang}   
        \hspace{-.1in} Zhang X., 1999, IAUS, 183, 276 

\bibitem[Zhang et al. 1997]{MIYUN}
        \hspace{-.1in} Zhang X., Zheng Y., Chen H., Wang S., Cao S., Peng B., Nan R., 1997, A\&AS, 121, 59

\end{thebibliography}
\end{document}